\documentclass[preprint,
normalem,
eqsecnum,
amsmath,amssymb,
amsmath,amssymb, aps,
]{revtex4}

\usepackage[dvipsnames]{xcolor}
\usepackage{epsfig}
\usepackage{amsfonts}
\usepackage{amsmath}
\usepackage{wasysym}
\usepackage{amssymb}
\usepackage{bm}
\usepackage{float}
\usepackage{enumitem}

\usepackage{graphicx}
\usepackage{ulem}

\usepackage[utf8]{inputenc}
\usepackage{bbm} 				
\usepackage{epstopdf}				
\usepackage{verbatim}    			
\usepackage{sidecap}                
\usepackage{indentfirst}

\newcommand{\be}{\begin{equation}}
\newcommand{\ee}{\end{equation}}
\newcommand{\bea}{\begin{eqnarray}}
\newcommand{\eea}{\end{eqnarray}}

\usepackage[dvipsnames]{xcolor}
\usepackage{graphicx}
\usepackage{dcolumn}
\usepackage{bm}

\usepackage[%
  colorlinks=true,
  urlcolor=blue,
  linkcolor=blue,
  citecolor=blue
]{hyperref}

\begin{document}

\bibliographystyle{apsrev4-2}


\title{Double Multiple-Relaxation-Time model of Lattice-Boltzmann \\
Magnetohydrodynamics at Low Magnetic Reynolds Numbers}

\author{B. Magacho$^{1}$\footnote{magacho@pos.if.ufrj.br}, H. S. Tavares$^1$, L. Moriconi$^1$, and
J. Loureiro$^2$}
\affiliation{$^1$Instituto de F\'\i sica, Universidade Federal do Rio de Janeiro,
C.P. 68528, CEP: 21941-972, Rio de Janeiro, RJ, Brazil,}
\affiliation{$^2$Programa de Engenharia Mecânica, Coordenação dos Programas de Pós-Graduação em Engenharia, 
Universidade Federal do Rio de Janeiro, \\ {C.P. 68503, CEP: 21945-970, Rio de Janeiro, RJ, Brazil}}




\begin{abstract}
We develop an improved lattice-Boltzmann numerical scheme to solve magnetohydrodynamic (MHD) equations in the regime of low magnetic Reynolds numbers, grounded on a manifestly Galilean covariant modeling of the Navier-Stokes equations. The simulation of the magnetic induction equation within the lattice-Boltzmann approach to MHD has been usually devised along the lines of the simplest phenomenological description, the single relaxation time (SRT) model. In order to deal with well-known stability difficulties of the SRT framework, we introduce, alternatively, a multi-relaxation-time technique for the solution of the magnetic induction equation, combined with a novel boundary condition method to cope with the subtleties of magnetic Boltzmann-like distributions on curved boundaries. As an application, we investigate open issues related to the description of transient flow regimes in MHD pipe flows, subject to non-uniform magnetic fields.
\end{abstract}

\maketitle


\section{\label{sec:level1}INTRODUCTION}

Magnetohydrodynamic (MHD) flows play a key role in a number of particularly important transport phenomena. The flow of electrically conducting fluids under the action of external electromagnetic fields is associated to remarkable dynamical effects in nuclear fusion, metallurgic casting setups, drag reduction, not to mention many other technologically relevant instances \cite{davidson1999,thomas2001,Marino2008,Tsinober}.

A perhaps more fundamental motivation for the study of MHD flows relies on the complex dynamics of coherent structures such as hairpin vortices and low-speed streaks that are generally observed in turbulent wall-bounded flows \cite{adrian2007,moriconi2009,dennis2014}.
The relaminarization of MHD turbulent flows by the magnetic dissipation of coherent structures 
and its detailed mechanisms are still the matter of open debate in the contemporary literature
\cite{Tsinober,lee_choi_2001,kobayashi2008,chaudhary_vanka_thomas2010,krasnov_zikanov_boeck_2012,zikanov_krasnov_boeck_thess_rossi,moriconi2020}.

Due to its practicality in dealing with boundary conditions (geometric or not) of different types, and its straightforward scalability in connection with the use of parallel computing strategies, the Lattice-Boltzmann Method (LBM) for the simulation of hydrodynamic equations \cite{Korner2006, Touil2016, Schornbaum2016, Xu_2018, LATT2021334} has received a great deal of attention along recent years. A main contributing factor for the growing interest in the LBM has been the considerable improvement of related hardware plataforms, based on very efficient - and by now more easily available - multi-threaded CPU and GPU devices.

The LBM, devised as a phase-space discretized version of the Boltzmann equation, is able to recover the usual Navier-Stokes equations of fluid  dynamics, from a mesoscopic modeling starting point \cite{chen_doolen1998,cercignanibook}. The LBM can also be extended, in principle, to simulate the more involved standard MHD equations, which consist of the electromagnetically forced Navier-Stokes equations coupled to the magnetic induction equation \cite{DELLAR2002, Croisille1995, Bouchut1999, PATTISON2008}. 
\vspace{0.0cm}

Lattice-Boltzmann simulations of MHD flows in the presence of walls and non-uniform external magnetic fields are, however, commonly hampered by numerical instabilities and the lack of systematic procedures for the implementation of boundary conditions. It is, in particular, of utmost interest to address a consistent solution of these issues in the regime of low magnetic Reynolds numbers, a dynamical condition which is observed to hold in most industrial situations \cite{zikanov_thess_1998, Knaepen2008,knaepen_kassinos_carati_2004}. In a related analysis \cite{saraiva_etal}, we have successfully approached the problem under the point of view of the recently proposed single-step version of the LBM \cite{delgado_etal,rosis_etal}. 

In this work, alternatively, we establish another efficient solution for the implementation of lattice-Boltzmann simulations in low magnetic Reynolds number MHD. The key technical aspects of our discussion are the multi-relaxation-time \cite{dhumieresarticle1992} and the central moment \cite{derosis_huang_coreixas2019} approaches to the LBM. To illustrate the accuracy of the proposed method, we work out, as a case study, the problem of MHD pipe flows under the presence of external static non-homogeneous magnetic fields.
\vspace{-0.0cm}

This paper is organized as follows. The essential points of the LBM are briefly reviewed in Sec.~\ref{LBM_METHOD_sec}. The implementation of the LBM in the MHD context is discussed in Sec.~\ref{Mag_field_sec}, where we detail the multi-relaxation-time modeling of the magnetic induction equation. In Sec.~\ref{bound_condi_sec}, we concentrate our attention on the treatment of boundary conditions in non-trivial geometries.
Numerical results, including a careful study of benchmark simulations, are presented in Sec.~\ref{sim_res_sec}. Finally, in Sec.~\ref{conc_sec}, we summarize our findings and point out directions of further research. 


\section{\label{LBM_METHOD_sec}The Lattice-Boltzmann Method}
The LBM in its general form is put forward as a two-step algorithm to solve a time-discretized Boltzmann evolution equation in a cubic lattice with unit lattice parameter. The lattice-Boltzmann equation reads as
\begin{equation}
    f_i(\boldsymbol{x} + \boldsymbol{c}_i,t+1) - f_i(\boldsymbol{x},t) = \boldsymbol{L}[f_i(\boldsymbol{x},t)] \ , \
    \label{generalLBM}
\end{equation}
where $f_i(\boldsymbol{x},t)$ represents the distribution of the ``molecular" population which has discretized velocity $\boldsymbol{c}_i$, while $\boldsymbol{L}$ is a collision operator, which can be defined in various phenomenologically motivated ways. The LHS and RHS of Eq. (\ref{generalLBM}) are, respectively, the so-called streaming and collision steps of the discretized Boltzmann equation. We stick, throughout this work, to the D3Q27 lattice formulation of hydrodynamics (three-dimensional cubic lattice; 27 discretized velocity vectors) of the LBM, where $i=0,1,...,26$ and the mesoscopic velocities $\boldsymbol{c}_i$ have their Cartesian components organized along three 27-dimensional column vectors,
\begin{eqnarray}
&& \vert c_{ix} \rangle =  (0, 1,-1, 0, 0,0 0, 1,-1, 1,-1, 1,-1, 1,-1, 0, 0,0, 0, 1,-1, 1,-1, 1,-1, 1,-1)^T  \ , \  \nonumber \\
    && \vert c_{iy} \rangle =  (0, 0, 0, 1,-1, 0, 0, 1, 1,-1,-1, 0, 0, 0, 0, 1,-1, 1,-1, 1, 1,-1,-1, 1, 1,-1,-1)^T \ , \ \nonumber \\
    && \vert c_{iz} \rangle = (0, 0, 0, 0, 0, 1,-1, 0, 0, 0, 0, 1, 1,-1,-1, 1, 1,-1,-1, 1, 1, 1, 1,-1,-1,-1,-1)^T \ . \  \nonumber \\
\end{eqnarray} 

Macroscopic quantities as the fluid density $\rho$ and the velocity field $\mathbf{u}$ are calculated by means of
\begin{eqnarray}
&&\rho = \sum_i f_i \ , \ \label{macroscopic_rho} \\
&&\mathbf{u} = \frac{1}{\rho}\sum_i \mathbf{c}_i f_i + \frac{\mathbf{F}}{2\rho}  \ , \
    \label{macroscopic_u}
\end{eqnarray}
where $\mathbf{F}$ stands for an arbitrary external force.



The simplest and by far the most popular collision model is the one proposed long ago by Bhatnagar, Gross, and Krook (BGK) \cite{BGK1954}, also known as the single relaxation time (SRT) model, where
\begin{equation}
    \boldsymbol{L}[f_i(\boldsymbol{x},t)] \equiv -\frac{1}{\tau}(f_i - f_i^{eq}) \ . \  \label{srt}
\end{equation}
In (\ref{srt}), $\tau$ defines a relaxation time and $f_i^{eq}$ is the discretized version of the Maxwell-Boltzmann equilibrium distribution associated to the molecular population which has
velocity $\boldsymbol{c}_i$. The distribution $f_i^{eq}$ depends not only on $\boldsymbol{c}_i$, but also 
on the velocity field $\mathbf{u}$ according to a well-defined prescription (to be made explicit below). Despite its simplicity and successful applications in the continuum Boltzmann equation setting, the SRT-LBM model can break down, as an artifact of lattice discretization, when further time scales come into play, as it is the case in hydrodynamic instabilities \cite{coveney_succi_dhumieres_ginzburg2002,hosseini2019}.


The crucial point, in order to improve the numerical convergence of the LBM, is to replace the SRT collision model by an alternative one which can address the otherwise missed physics of the related transport problem \cite{coreixas_chopard_latt2019,coreixas_wissocq_chopard_latt2020}. This goal is, to a great extent, accomplished by the multiple-relaxation-time (MRT) model \cite{dhumieresarticle1992}. Its main idea is to switch the focus from the usual space of populations,
\be
\vert f \rangle = (f_0,f_1,...,f_{26})^T \ , \ 
\ee
to a specifically defined space of velocity moments,
\be
\vert m \rangle \equiv (m_0,m_1,...,m_{26})^T  \ , \ 
\ee
where each one of the above individual moments is assigned, in principle, to independently tuned relaxation time scales. This procedure turns out to improve the numerical stability of the LBM, once physically important moments, as the ones that contribute to the Reynolds stress tensor, may relax to the local equilibrium, by construction, in a faster way than high-order and other non-physically relevant moments \cite{krugerbook,dhumieresarticle1992,lallemand_Luo2000}.

A problematic issue here, as discussed in \cite{malaspinas2015,Coreixas_wissocq_puigt_boussuge_sagaut2017,coreixas_phd_thesis,coreixas_chopard_latt2019}, is that the Boltzmann equilibrium distributions, $f_i^{eq}$, when usually truncated at second order in a power series expansions of the fluid velocity components, do not lead to Galilean invariant forcing terms in the D3Q27 lattice framework. It is possible, nevertheless, to overcome this difficulty, with the help of the following sixth-order expansions \cite{derosis2017,derosis_luo2019},
\begin{eqnarray}
            &&f_i^{eq} =  \omega_i \rho \Bigg  \{ 1 + \frac{\mathbf{c}_i \cdot \mathbf{u}}{c_s^2} + \frac{1}{2c_s^4} \Bigg [ \mathcal{H}_{ixx}^{(2)}u_x^2 + \mathcal{H}_{iyy}^{(2)}u_y^2 + \mathcal{H}_{izz}^{(2)}u_z^2 + 2 \Bigg ( \mathcal{H}_{ixy}^{(2)}u_x u_y + \mathcal{H}_{ixz}^{(2)}u_x u_z + \nonumber \\
            &+&\mathcal{H}_{iyz}^{(2)}u_y u_z \Bigg ) \Bigg ] + \frac{1}{2c_s^6} \Bigg [ \mathcal{H}_{ixxy}^{(3)}u_x^2 u_y + \mathcal{H}_{ixxz}^{(3)}u_x^2 u_z + \mathcal{H}_{ixyy}^{(3)}u_x u_y^2 + \mathcal{H}_{ixzz}^{(3)}u_x u_z^2 + \mathcal{H}_{iyzz}^{(3)}u_y u_z^2 +  \nonumber \\
            &+&\mathcal{H}_{iyyz}^{(3)}u_y^2 u_z +2 \mathcal{H}_{ixyz}^{(3)}u_x u_y u_z \Bigg ] + \frac{1}{4c_s^8} \Bigg [ \mathcal{H}_{ixxyy}^{(4)}u_x^2 u_y^2 + \mathcal{H}_{ixxzz}^{(4)}u_x^2 u_z^2 + \mathcal{H}_{iyyzz}^{(4)}u_y^2 u_z^2 + \nonumber \\
            &+&2 \Bigg( \mathcal{H}_{ixyzz}^{(4)}u_x u_y u_z^2 + \mathcal{H}_{ixyyz}^{(4)}u_x u_y^2 u_z + \mathcal{H}_{ixxyz}^{(4)}u_x^2 u_y u_z \Bigg ) \Bigg ] + \frac{1}{4c_s^{10}} \Bigg [ \mathcal{H}_{ixxyzz}^{(5)}u_x^2 u_y u_z^2 + \nonumber \\ 
            &+& \mathcal{H}_{ixxyyz}^{(5)}u_x^2 u_y^2 u_z + \mathcal{H}_{ixyyzz}^{(5)}u_x u_y^2 u_z^2 \Bigg ] + \frac{1}{8c_s^{12}}\mathcal{H}_{ixxyyzz}^{(6)}u_x^2 u_y^2 u_z^2  \Bigg \} \ , \ 
            \label{BoltzmannEquilibriumDistribution}
\end{eqnarray}
where $c_s$ is the sound velocity, $\mathcal{H}_{i}^{(n)}$ denotes a tensor Hermite polynomial of order $n \leq 6$, and $\omega_i$ are lattice-Boltzmann weights, defined by 
\begin{equation}
\omega_1 =...= \omega_6 = \omega_0/4 \ , \ \omega_7 =...= \omega_{18} = \omega_0/16 \ , \ \omega_{19} =...= \omega_{26} = \omega_0 / 64 \ , \ \label{omegas}
\end{equation}
with $\omega_0 = 8/27$.


The central moments (CM) model -- also referred to as the cascaded lattice-Boltzmann model \cite{derosis_huang_coreixas2019} -- is a further refinement of the MRT approach, which improves the stability of lattice-Boltzmann simulations, from the analysis of Galilean invariant moments \cite{Nie2008}. In order to introduce a collision operator with central moments, one should shift the original lattice velocities of the MRT model, defined in the ``laboratory frame", to a set of velocities measured in the local comoving reference frame attached to the fluid elements \cite{geier2006}, viz.,
\be
\overline{c}_{ix} =  c_{ix} - u_x  \ , \  
\overline{c}_{iy}  =  c_{iy} - u_y \ , \  
\overline{c}_{iz} = c_{iz} - u_z  \ . \ \label{cm}
\ee
The central moments are now defined as the inner products
\begin{equation}
k_i \equiv \langle T_i \vert f \rangle \ , \ 
k^{eq}_i \equiv \langle T_i \vert f^{eq} \rangle \ , \ \label{kieq}
\end{equation}
where \cite{derosis2017},
\begin{eqnarray}
&& \vert T_{0} \rangle = \vert 1,...,1 \rangle \ , \ 
 \vert T_{1} \rangle = \vert \overline{c}_{ix} \rangle \ , \  
 \vert T_{2} \rangle = \vert \overline{c}_{iy} \rangle \ , \ 
 \vert T_{3} \rangle = \vert \overline{c}_{iz} \rangle \ , \ 
\vert T_{4} \rangle = \vert \overline{c}_{ix}\overline{c}_{iy} \rangle \ , \  \nonumber \\
&& \vert T_{5} \rangle = \vert \overline{c}_{ix}\overline{c}_{iz} \rangle \ , \ 
\vert T_{6} \rangle = \vert \overline{c}_{iy}\overline{c}_{iz} \rangle \ , \ 
\vert T_{7} \rangle = \vert \overline{c}_{ix}^2 - \overline{c}_{iy}^2 \rangle \ , \ 
\vert T_{8} \rangle = \vert \overline{c}_{ix}^2 - \overline{c}_{iz}^2 \rangle \ , \ \nonumber \\
&& \vert T_{9} \rangle = \vert \overline{c}_{ix}^2 + \overline{c}_{iy}^2 + \overline{c}_{iz}^2 \rangle \ , \  \vert T_{10} \rangle = \vert \overline{c}_{ix} \overline{c}_{iy}^2 + \overline{c}_{ix} \overline{c}_{iz}^2 \rangle \ , \ \vert T_{11} \rangle = \vert \overline{c}_{ix}^2 \overline{c}_{iy} + \overline{c}_{iy} \overline{c}_{iz}^2 \rangle \ , \  \nonumber \\
&& \vert T_{12} \rangle = \vert \overline{c}_{ix}^2\overline{c}_{iz} + \overline{c}_{iy}^2\overline{c}_{iz} \rangle \ , \   \vert T_{13} \rangle = \vert \overline{c}_{ix} \overline{c}_{iy}^2 - \overline{c}_{ix} \overline{c}_{iz}^2 \rangle \ , \  \vert T_{14} \rangle = \vert \overline{c}_{ix}^2 \overline{c}_{iy} - \overline{c}_{iy} \overline{c}_{iz}^2 \rangle \ , \  \nonumber \\
&& \vert T_{15} \rangle = \vert \overline{c}_{ix}^2\overline{c}_{iz} - \overline{c}_{iy}^2\overline{c}_{iz} \rangle \ , \ \vert T_{16} \rangle = \vert \overline{c}_{ix} \overline{c}_{iy} \overline{c}_{iz} \rangle \ , \  \vert T_{17} \rangle = \vert \overline{c}_{ix}^2 \overline{c}_{iy}^2 + \overline{c}_{ix}^2 \overline{c}_{iz}^2 + \overline{c}_{iy}^2 \overline{c}_{iz}^2 \rangle \ , \  \nonumber \\
&& \vert T_{18} \rangle = \vert \overline{c}_{ix}^2 \overline{c}_{iy}^2 + \overline{c}_{ix}^2 \overline{c}_{iz}^2 - \overline{c}_{iy}^2 \overline{c}_{iz}^2 \rangle \ , \   \vert T_{19} \rangle = \vert \overline{c}_{ix}^2 \overline{c}_{iy}^2 - \overline{c}_{ix}^2\overline{c}_{iz}^2 \rangle \ , \  \vert T_{20} \rangle = \vert \overline{c}_{ix}^2 \overline{c}_{iy} \overline{c}_{iz}  \rangle \ , \  \nonumber \\
&& \vert T_{21} \rangle = \vert \overline{c}_{ix} \overline{c}_{iy}^2 \overline{c}_{iz} \rangle \ , \   \vert T_{22} \rangle = \vert \overline{c}_{ix} \overline{c}_{iy} \overline{c}_{iz}^2 \rangle \ , \   \vert T_{23} \rangle = \vert \overline{c}_{ix} \overline{c}_{iy}^2 \overline{c}_{iz}^2  \rangle \ , \  \vert T_{24} \rangle = \vert \overline{c}_{ix}^2 \overline{c}_{iy} \overline{c}_{iz}^2 \rangle \ , \  \nonumber \\
&& \vert T_{25} \rangle = \vert \overline{c}_{ix}^2 \overline{c}_{iy}^2 \overline{c}_{iz} \rangle \ , \  \vert T_{26} \rangle = \vert \overline{c}_{ix}^2 \overline{c}_{iy}^2 \overline{c}_{iz}^2  \rangle \ . \
\label{Ts}
\end{eqnarray}
We notice, as a trivial remark, that if {\hbox{$u_x=u_y=u_z=0$}} in (\ref{cm}), then $k_i$ is just the standard moment $m_i$ of the MRT formalism. The post-collision central moment vector is given by
\begin{equation}
    \vert k^* \rangle = (\mathbf{I} - \mathbf{\Lambda})\vert k \rangle + \mathbf{\Lambda} \vert k^{eq} \rangle + \Bigg ( \mathbf{I} - \frac{\mathbf{\Lambda}}{2} \Bigg )\vert R \rangle \ , \ \label{k*}
\end{equation}
where the $i$-th component of $| R \rangle$ is
\be
R_i = \langle T_i \vert \mathcal{F} \rangle \ , \ \label{Ri}
\ee
and $|\mathcal{F} \rangle$ stands for the equilibrium contribution 
derived from the Boltzmann equation that includes an arbitrary 
external force {\bf{F}}. We have, in general \cite{guo2002},
\begin{equation}
   | \mathcal{F} \rangle = -\mathbf{F} \cdot \nabla_{\bf{c}} | f^{eq} \rangle \ . \ 
    \label{Forcing_term}
\end{equation}
The matrix elements of the $27 \times 27$ collision matrix $\mathbf{\Lambda}$ introduced in (\ref{k*}) are defined as
\begin{equation}
\Lambda_{ij} =
    \begin{cases}
      \delta_{ij} \ , \ \text{if $i \in \{ 5,6,7,8,9 \} $} \ , \ \\
    \omega \delta_{ij} \ , \ \text{otherwise} \ , \ \\
    \end{cases}    \label{lambda}
\end{equation}
where \textcolor{black}{$\omega = 1/\tau = (\nu/c_s^2 + 0.5)^{-1} >  1$ with $\nu$ being the kinematic viscosity}. Taking a look at the organization of moments in (\ref{Ts}), we see, from (\ref{lambda}), that only the second order moments associated to the stress tensor approach local equilibirum with relaxation times $1 / \omega < 1$, while the other moments, slower and not physically relevant, are expected
not to produce spurious effecs on the dynamical evolution of the Boltzmann populations \cite{YOSHIDA2010,leveque_hal,coveney_succi_dhumieres_ginzburg2002,hosseini2019}.

Working with the set of Eqs.~(\ref{BoltzmannEquilibriumDistribution}-\ref{lambda}), we find the post-collision moments,
\begin{eqnarray}
&& k_0^* = \rho \ , \  k_1^* = F_x/2 \ , \  k_2^* = F_y/2 \ , \  k_3^* = F_z/2 \ , \ k_4^* = (1-\omega)k_4 \ , \  k_5^* = (1-\omega)k_5 \ , \ \nonumber \\
&& k_6^* = (1-\omega)k_6 \ , \  k_7^* = (1-\omega)k_7 \ , \  k_8^* = (1-\omega)k_8 \ , \  k_9^* = 3\rho c_s^2 \ , \ k_{10}^* = F_x c_s^2 \ , \  \nonumber \\
&& k_{11}^* = F_y c_s^2 \ , \  k_{12}^* = F_z c_s^2 \ , \ k_{13}^* =  k_{14}^* = k_{15}^* =  k_{16}^* = 0 \ , \  k_{17}^* = \rho c_s^2 \ , \ k_{18}^* = \rho c_s^4 \ , \ 
\nonumber  \\
&& k_{19}^* = k_{20}^* =  k_{21}^* = k_{22}^* = 0 \ , \ k_{23}^* = F_x c_s^4/2 \ , \ 
k_{24}^* = F_y c_s^4/2 \ , \  
k_{25}^* = F_z c_s^4/2 \ , \ \nonumber \\
&& k_{26}^* = \rho c_s^6 \ , \
\end{eqnarray}
where
\begin{eqnarray}
&& k_4 = \sum_i f_i \overline{c}_{ix}\overline{c}_{iy} \ ,  \  k_5 = \sum_i f_i \overline{c}_{ix}\overline{c}_{iz} \ , \  k_6 = \sum_i f_i \overline{c}_{iy}\overline{c}_{iz} \ , \ \nonumber \\
&&k_7 = \sum_i f_i (\overline{c}_{ix}^2 - \overline{c}_{iy}^2) \ , \  k_8 = \sum_i f_i (\overline{c}_{ix}^2 - \overline{c}_{iz}^2) \ . \
\end{eqnarray}
Note that the $k^*$'s are all Galilean invariant quantities. The post-collision populations are then obtained by means of the inverse transformation,
\begin{equation}
\vert f^* \rangle = T^{-1}\vert k^* \rangle \ , \ \label{k*f*}
\end{equation}
with $T$ being a 27$\times$27 matrix defined by the matrix elements
\be
T_{ij} \equiv \langle T_i | j \rangle \  , \ 
\ee
where
\be
| j \rangle  = (\delta_{0j}, \delta_{1j},..., \delta_{26j})^T \ . \ 
\ee
Once we are done with the evaluation of $|k^* \rangle$ from the application of Eq. (\ref{k*}), and we have computed $| f^* \rangle$ with the help
of Eq. (\ref{k*f*}), the next recursive step for updating the Boltzmann distributions, as already discussed in the presentation of (\ref{generalLBM}), is to perform the simpler streaming step over the lattice sites.

We have so far recalled the essential ingredients of the MRT/CM setup of the LBM for pure hydrodynamics. We are now ready to address analogous considerations for the problem of magnetohydrodynamic flows.

\section{\label{Mag_field_sec}LBM for Magnetohydrodynamics}

The set of magnetohydrodynamic equations for incompressible conducting fluids can be written down as the Lorentz forced Navier-Stokes equations combined with the induction equation for the magnetic field dynamics, as derived from Faraday's and Ohm's laws \cite{Biskampbook,shercliffbook}. More concretely,
\begin{eqnarray}
&&\partial_t \mathbf{u} = -\nabla (p/\rho) - (\mathbf{u} \cdot \mathbf{\nabla})\mathbf{u} + \nu \nabla^2 \mathbf{u} + \frac{\mathbf{F}_m}{ \rho} \ , \ 
    \label{NSMagForce} \\
&&\partial_t \mathbf{B} =  - (\mathbf{u} \cdot \mathbf{\nabla})\mathbf{B} + (\mathbf{B} \cdot \mathbf{\nabla})\mathbf{u} + \eta \nabla^2 \mathbf{B} \ , \  \label{completeIE} \\
&&\mathbf{\nabla} \cdot \mathbf{u} = \mathbf{\nabla} \cdot \mathbf{B} = 0 \ , \ 
\label{divV}
\end{eqnarray}
where $\mathbf{B}$ stands for the total magnetic field (external plus induced), $\eta$ is the magnetic diffusivity, and $\mathbf{F}_m$ is the Lorentz force, defined as
\begin{equation}
    \mathbf{F}_m = \frac{1}{\mu}(\nabla \times \mathbf{B}) \times \mathbf{B} \ . \ \label{lforce}
\end{equation}
Here $\mu$ is the magnetic permeability, which is related to the magnetic diffusivity and the electric conductivity $\sigma$ as $\eta = 1/(\mu \sigma)$.
For a proper lattice realization of the above differential operators, we follow the prescriptions established in \cite{THAMPI2013}. In this way, anisotropic effects of discretization are mitigated and the only sources of anisotropy in hydrodynamic scales are expected to come either from boundary conditions (geometric or not) or from external fields.

Of course, it is possible to solve Eq. \eqref{completeIE} without resorting to the LBM, at the expense, in general, of Fourier non-locality, besides more involved and costly treatments of boundary conditions. It is clear, on the other hand, that a complete LBM simulation of the velocity and magnetic field dynamics should be necessarily related to two collision models: one for Eq. (\ref{NSMagForce}) and the other one for (\ref{completeIE}). As it was discussed in the previous section, Eq. (\ref{NSMagForce}) can be numerically solved along the lines of the LBM in a number of different ways. 

The LBM account of the magnetic induction equation (\ref{completeIE}), on its turn, was addressed in the seminal work of Dellar \cite{DELLAR2002}, who derived a BGK model for the evolution of magnetic vector valued distributions, unifying a MHD kinetic approach with a general construction of BGK collision models for a variety of systems \cite{Croisille1995,Bouchut1999}. \textcolor{black}{This double collision BGK-BGK model for the LBM simulation of MHD flows has been applied to different problems, such as laminar MHD channel flows and the three-dimensional Orszag-Tang vortex, among other examples \cite{DELLAR2002}. In subsequent works, Riley et al. \cite{Riley2007,PATTISON2008} and de Rosis et al. \cite{derosis_huang_coreixas2019,leveque_hal} established, respectively, applications using the MRT-BGK to nuclear fusion and CM-BGK to homogeneous and isotropic MHD turbulence, that were alternatives to the original BGK-BGK formulation. }
 
Noticing that there is actually room for variations of the LBM approach to the magnetic induction equation (\ref{completeIE}), our aim in this work, in short and objective words, is to close a gap in the literature and develop a CM-MRT collision model for the solution of the coupled Eqs. (\ref{NSMagForce} - \ref{completeIE}). As it will become clear in Sec. III, the CM-MRT approach not only improves the accuracy of solutions for the velocity and magnetic fields, but also fix the eventual numerical instabilities associated to the BGK-BGK model, when spurious transient modes are not properly suppressed.
\vspace{0.3cm}

\leftline{\it{The Magnetic BGK Collision Model}}
\vspace{0.3cm}

It is interesting, before addressing the CM-MRT framework, to summarize the Dellar's BGK model for the magnetic induction equation. The magnetic vector field is represented as the 0-th order moment of vector valued distributions $\boldsymbol{g}_i$, which in a D3Q7 formulation, reads as
\begin{equation}
    \mathbf{B} = \sum_{i=0}^{6}\boldsymbol{g}_i \ , \ \label{Bg}
\end{equation}
with $\boldsymbol{g}_i (\boldsymbol{x},t) \equiv \boldsymbol{g}(\boldsymbol{\xi}_i;\boldsymbol{x},t)$, where
\bea
&&\boldsymbol{\xi}_0 = (0,0,0) \ , \ 
\boldsymbol{\xi}_1 = (1,0,0) \ , \ 
\boldsymbol{\xi}_2 = (-1,0,0) \ , \ 
\boldsymbol{\xi}_3 = (0,1,0) \ , \ \nonumber \\
&&\boldsymbol{\xi}_4 = (0,-1,0) \ , \ 
\boldsymbol{\xi}_5 = (0,0,1) \ , \ {\hbox{and }} \boldsymbol{\xi}_6 = (0,0,-1) 
\eea
are the discretized magnetic vector ``velocities" defined in the three-dimensional 
cubic lattice.

The discretized time evolution of the vector-valued distributions $\boldsymbol{g}_i$ is carried out through the usual two-step collision-streaming iteration, applied to the BGK modeling equation, viz.,
\begin{equation}
\boldsymbol{g}_i(\boldsymbol{x} + \boldsymbol{\xi}_i, t + 1) - \boldsymbol{g}_i(\boldsymbol{x}, t) = - \frac{1}{\tau_m}[\boldsymbol{g}_i(\boldsymbol{x}, t) - \boldsymbol{g}^{eq}_i(\boldsymbol{x}, t)] \ . \ 
\label{bgkmagnetic}
\end{equation}
Above, $\tau_m$ is the magnetic relaxation time which depends on the magnetic diffusivity $\eta$ as
\begin{equation}
    \tau_m = c_m^{-2} \eta + \frac{1}{2} \ , \ \label{taum}
\end{equation}
where $c_m$ is a magnetic velocity parameter, analogous to the sound velocity $c_s$. The magnetic equilibrium distributions $\boldsymbol{g}^{eq}_i$ are given by
\begin{equation}{\label{gEq}}
\boldsymbol{g}^{eq}_i = 
w_i \left \{ \mathbf{B} + c_m^{-2} [ ( \boldsymbol{\xi}_i \cdot \mathbf{u}) \mathbf{B}-   ( \boldsymbol{\xi}_i \cdot \mathbf{B}) \mathbf{u} ] \right \} \ , \
\end{equation}
where
\be
w_0 = 1/4 \ , \  w_1 = ... = w_6 = 1/8  
\ee
are magnetic lattice weights, analogous to the ones given in (\ref{omegas}).
\vspace{0.3cm}



\leftline{\it{The Magnetic MRT Collision Model}}
\vspace{0.3cm}

We note that Eq. \eqref{completeIE} can be interpreted as a diffusion equation for the advected magnetic field, additionally perturbed by a
source contribution", given by the second term on its RHS. Our attention, thus, is driven to the previously developed MRT collision models for thermal diffusion equations \cite{YOSHIDA2010}, which have an analogous formal structure. They can be, in fact, adapted to the magnetic field problem as detailed below.

Define, to start, the linear map $M$ from the vector-valued distributions, $\boldsymbol{g}_i$, to the vector-valued moment distributions $\boldsymbol{m}_i$,
\begin{equation}
    \boldsymbol{m}_i  = \langle M_i \vert \boldsymbol{g} \rangle \ , \ \label{mi}
\end{equation}
where $\vert \boldsymbol{g} \rangle = (\boldsymbol{g}_0, \boldsymbol{g}_1,...,\boldsymbol{g}_6)^T$, and
\begin{equation}
    M = 
    \begin{bmatrix}
    \langle 1 \vert \\
    \langle \xi_x \vert \\
    \langle \xi_y \vert \\
    \langle \xi_z \vert \\
    \langle 6 - 7\boldsymbol{\xi}^2 \vert \\
    \langle 3 \xi_x^2 - \boldsymbol{\xi}^2 \vert \\
    \langle \xi_y^2 - \xi_z^2 \vert 
    \end{bmatrix}
    =
    \begin{bmatrix}
    1 & 1 & 1 & 1 & 1 & 1 & 1\\
    0 & 1 & -1 & 0 & 0 & 0 & 0\\
    0 & 0 & 0 & 1 & -1 & 0 & 0\\
    0 & 0 & 0 & 0 & 0 & 1 & -1\\
    6 & -1 & -1 & -1 & -1 & -1 & -1\\
    0 & 2 & 2 & -1 & -1 & -1 & -1\\
    0 & 0 & 0 & 1 & 1 & -1 & -1
    \end{bmatrix} \ . \
\end{equation}
Eq. \eqref{bgkmagnetic} is now replaced by 
\begin{equation}
\boldsymbol{g}_i(\mathbf{x} + \boldsymbol{\xi}_i, t + 1) - \boldsymbol{g}_i(\mathbf{x}, t)=- \sum_{j=0}^6 (M^{-1} S M)_{ij} 
[ \boldsymbol{g}_j(\mathbf{x}, t) - \boldsymbol{g}^{eq}_j(\mathbf{x, t}) ] \ , \
\label{premrtmag}
\end{equation}
where the MRT collision matrix $S$, is defined as
\begin{equation}
    S^{-1} = 
    \begin{bmatrix}
    \tau_0 & 0 & 0 & 0 & 0 & 0 & 0\\
    0 & \tau_{xx} & \tau_{xy} & \tau_{xz} & 0 & 0 & 0\\
    0 & \tau_{yx} & \tau_{yy} & \tau_{yz} & 0 & 0 & 0\\
    0 & \tau_{zx} & \tau_{zy} & \tau_{zz} & 0 & 0 & 0\\
    0 & 0 & 0 & 0 & \tau_{4} & 0 & 0\\
    0 & 0 & 0 & 0 & 0 & \tau_{5} & 0\\
    0 & 0 & 0 & 0 & 0 & 0 & \tau_{6}
    \end{bmatrix} \label{matrix} \ . \
\end{equation}
The off-diagonal elements of (\ref{matrix}) can be used to model anisotropic diffusion \cite{YOSHIDA2010}. We restrict ourselves, however, to the case of isotropic magnetic diffusion,
\begin{equation}
\tau_{xx} =  \tau_{yy} =  \tau_{zz} = \tau_m \ , \ \tau_{xy} = \tau_{yx} =\tau_{xz} = \tau_{zx} = \tau_{yz} =\tau_{zy} = 0 \ , \ \label{taus}
\end{equation}
where $\tau_m$ is the usual collision relaxation time, and, by convention,
\be
\tau_0 = \tau_4 =\tau_5 = \tau_6 = 1 \ . \
\ee
Since we are particularly interested in the case of low magnetic Reynolds numbers, when the magnetic diffusivity is large, we will work with $\tau_m > 1$, as it can be hinted by (\ref{taum}). The faster decaying moments, all associated to the same unit time scale, ensure, in an efficient way, that the magnetic field perturbations produced by the induced currents are quickly suppressed during the dynamical evolution. Note that the standard magnetic BGK collision model is recovered for the specific case $\tau_0 = \tau_4 = \tau_5 = \tau_6 = \tau_m$.


Applying, now, $M$ on both sides of (\ref{premrtmag}), we obtain, from  (\ref{mi}), the evolution equation for the magnetic moments,
\begin{equation}
\boldsymbol{m}_i (\boldsymbol{x} + \boldsymbol{\xi_i}, t + 1)  
- \boldsymbol{m}_i (\boldsymbol{x}, t)  =  - \sum_{j=0}^6 S_{ij}  [ \boldsymbol{m}_j (\boldsymbol{x}, t) - \boldsymbol{m}^{eq}_j  (\boldsymbol{x}, t) ]
\ . \  \label{moments}
\end{equation}
The multiplet of equilibrium moments, $\boldsymbol{m}_j^{eq}$, straightforwardly computed from Eq.~(\ref{gEq}),
is
\begin{equation}
\vert \boldsymbol{m}^{eq} \rangle = ( \mathbf{B} \ , \ u_x \mathbf{B} - B_x \mathbf{u} \ , \ 
u_y \mathbf{B} - B_y \mathbf{u} \ , \  u_z \mathbf{B} - B_z \mathbf{u} \ , \ 3\mathbf{B}/4 \ , \ 0 \ , \ 0 )^T \ . \ \label{meqB}
\end{equation}
After each full iteration cycle (collision $\&$ streaming) of the lattice-Boltzmann 
{\hbox{Eq. (\ref{moments})}},
the set of vector-valued distributions $\boldsymbol{g}_i$ is obtained by inverting (\ref{mi}). The 
magnetic field can then be computed over all the lattice sites through Eq. (\ref{Bg}).
\vspace{0.3cm}

\leftline{\it{The Quasi-Static Approximation}}
\vspace{0.3cm}

The quasi-static approximation \cite{zikanov_thess_1998, Knaepen2008,knaepen_kassinos_carati_2004} of the MHD Eqs. (\ref{NSMagForce}-\ref{divV}), when the magnetic backreaction from induced currents is negligible in comparison with the external magnetic field, is meaningful in the limit of small magnetic Reynolds and Prandtl numbers, given, respectively, by
\begin{equation}
    Re_m = \frac{uL}{\eta} {\hbox{ and }}  Pr_m = \frac{\nu}{\eta} = \frac{Re_m}{Re} \ . \
\end{equation}
We may write the total magnetic field $\mathbf{B}$ as the superposition of an external field $\mathbf{B}_{ext}$ and a relatively small fluctuating contribution, $\mathbf{b}$. That is,
\begin{equation}
\mathbf{B} = \mathbf{B}_{ext} + \mathbf{b} \ . \
\end{equation}
The quasi-static approximation means that 
\be
\mathcal{O}(\partial_t \mathbf{b}) \ll  \mathcal{O}(\partial_t \mathbf{B}_{ext}) 
\ee
and that the dynamo and magnetic advection effects in Eq. (\ref{completeIE}) are overcome by magnetic diffusion
\cite{zikanov_thess_1998,knaepen_kassinos_carati_2004}. More concretely,
\begin{equation}
\mathcal{O}((\mathbf{u} \cdot \nabla)\mathbf{b}) \approx \mathcal{O}((\mathbf{b} \cdot \nabla)\mathbf{u}) \ll \mathcal{O}(\eta \nabla^2 \mathbf{b}) \ . \ 
\end{equation}
By neglecting time derivatives of $\mathbf{b}$, Eq. (\ref{completeIE}) becomes, in the situation of a static external magnetic field,
\begin{equation}
     \eta \nabla^2 \mathbf{b} = (\mathbf{u} \cdot \mathbf{\nabla})\mathbf{B}_{ext} - (\mathbf{B}_{ext} \cdot \mathbf{\nabla})\mathbf{u} - \eta \nabla^2 \mathbf{B}_{ext}\ , \
     \label{Poisson_magnetic}
\end{equation}
where, of course, $\nabla \cdot \mathbf{b} = 0$.
In principle, thus, all that one needs to do here is to solve the Poisson Eq.~\eqref{Poisson_magnetic} for the fluctuating field $\mathbf{b}$. This is actually the usual practice in direct numerical simulations of the quasi-static MHD equations \cite{krasnov_zikanov_boeck_2012,liu_vanka_thomas2014}.

In a full lattice-Boltzmann simulation of Eqs. (\ref{NSMagForce}-\ref{divV}), however, 
the time derivative of $\mathbf{b}$ is not neglected. Eq. (\ref{completeIE}) is rewritten as
\begin{equation}
    \partial_t \mathbf{b} =  - (\mathbf{u} \cdot \mathbf{\nabla})\mathbf{B}_{ext} + (\mathbf{B}_{ext} \cdot \mathbf{\nabla})\mathbf{u} + \eta \nabla^2 (\mathbf{b} + \mathbf{B}_{ext}) \ . \
    \label{IELBM}
\end{equation}
{\it{Mutatis mutandis}}, the MRT modeling of (\ref{IELBM}) can be addressed here in the same way as previously encoded in Eq. (\ref{moments}). Eq. (\ref{meqB}) is now replaced by
\begin{equation}
    \vert \boldsymbol{m}^{eq} \rangle = ( \mathbf{b} \ , \ u_x \mathbf{B}_{ext} - B_{x,ext} \mathbf{u} \ , \  u_y \mathbf{B}_{ext} - B_{y,ext} \mathbf{u} \ , \ u_z \mathbf{B}_{ext} - B_{z,ext} \mathbf{u} \ , \ 3\mathbf{b}/4 \ , \ 0 \ , \ 0)^T.
\end{equation}
The quasi-static approximation for (\ref{IELBM}) is achieved in the lattice-Boltzmann language, as already outlined, through the specific choice $\tau_m > 1$ for the magnetic relaxation time.

The lattice-Boltzmann methodology here developed leads to great improvement in the specific case of flows bounded by insulating walls, where Dirichlet boundary conditions are imposed ($\mathbf{b} \vert_{walls} = 0$). The treatment of boundary conditions grows in complexity when bounding surfaces are not aligned with the simulation grid, a point we will emphasize in the numerical experiments reported in Sec. V.
\vspace{0.3cm}

\leftline{\it{Alternative Modeling of the Lorentz Force}}
\vspace{0.3cm}


The Lorentz force (\ref{lforce}), like any arbitrary external force, can be straightforwardly accounted in the LBM by means of the simulational scheme given by Eqs. (\ref{k*}-\ref{Forcing_term}). Note, however, that the evaluation of the spatial derivatives of the magnetic field in (\ref{lforce}) is likely to impose demands on the grid resolution, mainly in the case of turbulent flows, where one may expect the occurrence of faster fluctuations of the induced magnetic field over the lattice. There is, fortunately, a way to avoid the evaluation of derivatives of the magnetic field, as discussed in \cite{DELLAR2002}. By changing the equilibrium distributions to account for the Maxwell stress tensor in the first order moments, everything boils down to the replacement of the original equilibrium distributions $f_i^{eq}$, defined in (\ref{BoltzmannEquilibriumDistribution}), by additively corrected ones, $f_i^{eq} + f_{i,mag}^{eq}$, where
\begin{equation}
f_{i,mag}^{eq} = \frac{\omega_i}{2 c_s^4} \Bigg [ \frac{\vert \mathbf{c}_i \vert^2 \vert \mathbf{B} \vert^2}{D} - (\mathbf{c}_i \cdot \mathbf{B})^2  \Bigg ] \ . \ 
\end{equation}
Above, $D=2$ or 3 denotes the space dimension. The modified equilibrium distribution is then used to input (\ref{kieq}) into Eq. (\ref{k*}). This implies, in practical terms, 
that a contribution $\mathbf{\Lambda} T |f_{mag}^{eq} \rangle$ is added to the RHS of Eq. (\ref{k*}).  

\section{\label{bound_condi_sec} BOUNDARY CONDITIONS}

The introduction of geometric boundary conditions (BC) in the LBM is a point of further (and far from trivial) discussion for both the distributions $f_i$ and $\boldsymbol{g}_i$. The no-slip BC can systematically implemented for the velocity field by means of the Bouzidi technique \cite{bouzidi2001}, where, for non-Cartesian boundaries, $f_i$ depends in a well-defined way on the distance of the lattice nodes to the solid surfaces and boundary nodes. In the MHD context, Pattison et al. \cite{PATTISON2008} have introduced an extrapolation methodology to work with non-Cartesian boundaries, which, however, is likely to miss the accuracy gain that would be eventually obtained by a Bouzidi-like treatment of BCs. This issue naturally suggests an improvement in the implementation of magnetic BCs, which we discuss in the following.

Let $\boldsymbol{x}_f, \boldsymbol{x}_w$ and $\boldsymbol{x}_b$ represent the fluid, wall and boundary nodes, respectively (i.e., $\boldsymbol{x}_f$ and $\boldsymbol{x}_w$ are the nodes, inside and outside the flow region, respectively, which are placed in a minimal neighborhood of the boundary point $\boldsymbol{x}_b$). Consider, now, the distance ratio parameter
\begin{equation}
    \Delta \equiv \frac{\vert \boldsymbol{x}_f - \boldsymbol{x}_w \vert}{\vert \boldsymbol{x}_f - \boldsymbol{x}_b \vert} \ . \ \label{Delta}
\end{equation}
Eq. (\ref{Delta}) is, to be more precise, a bookkeeping definition of $\Delta$, to the extent that it
can depend on the lattice-Boltzmann directions $\boldsymbol{c}_i$ \cite{PATTISON2008,bouzidi2001,li_mei_klausner2013}.
The usual Bouzidi BCs \cite{bouzidi2001} are based on a interpolated version of the bounce-back scheme \cite{Ziegler1993, Ginzbourg1994}, as
\begin{eqnarray}
&&f_{\overline{i}}(\boldsymbol{x}_f,t+1) = \frac{1}{2\Delta}\tilde{f}_{i}(\boldsymbol{x}_f,t) + \frac{2\Delta - 1}{2\Delta}\tilde{f}_{\overline{i}}(\boldsymbol{x}_f,t)  \mbox{   for  } \Delta \geq \frac{1}{2} \ , \  \label{fbc1}\\
&&f_{\overline{i}}(\boldsymbol{x}_f,t+1) = 2\Delta \tilde{f}_{i}(\boldsymbol{x}_f,t) + (1 - 2\Delta) \tilde{f}_{i}(\boldsymbol{x}_{ff},t)  \mbox{   for  } \Delta < \frac{1}{2} \ , \ \label{fbc2}
\end{eqnarray}
where $\overline{i}$ denotes the direction opposite to \textcolor{black}{$\boldsymbol{c_i}$ (that is, - $\boldsymbol{c_i}$)}, $\tilde{f}_i$ represents the post-collision distribution, and $\boldsymbol{x}_{ff} \equiv \boldsymbol{x}_{f} - \mathbf{c}_{i}$. 

\begin{figure}[t]
    \centering
 \includegraphics[width = 10cm]{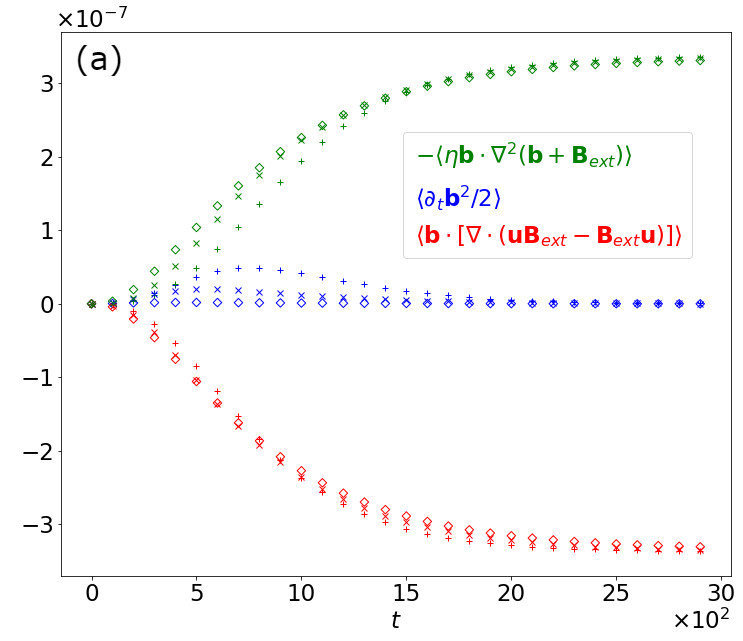}
   \includegraphics[width = 10cm]{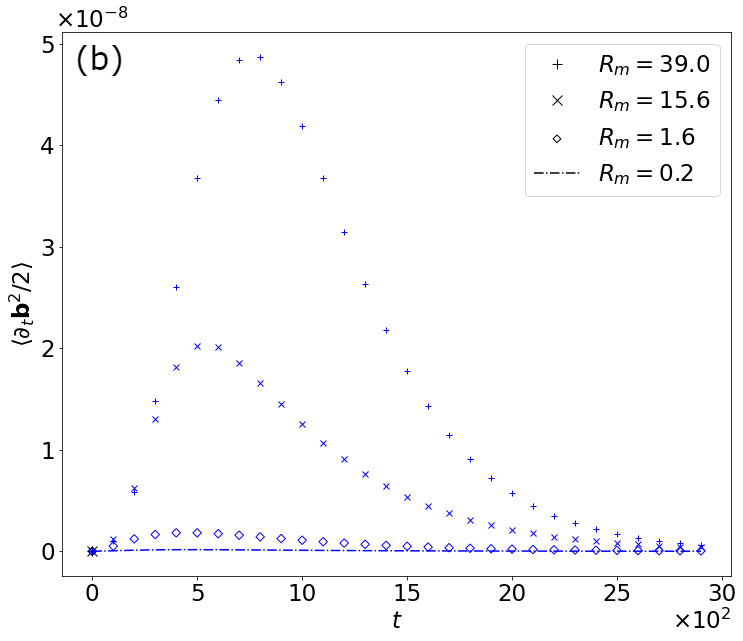}
    \caption{Results of the lattice-Boltzmann simulations at Ha = 5. (a) Production terms for the magnetic energy balance at $R_m = 39$ ($+$), $R_m = 15.6$ ($\times$), and $R_m = 1.6$ ($\diamond$). (b) A closer look at the time evolution of $\partial_t \mathbf{b}^2/2$ for various $R_m$.}
\end{figure}

To establish definitions of magnetic BCs which are analogous to (\ref{fbc1}) and (\ref{fbc2}), we, again, take advantage of the similarity that there is between the LBM approach to thermal transport and MHD.  Motivated by the modeling strategy of Li et al. \cite{li_mei_klausner2013}, who have focused on the problem of BC implementation in the thermal LBM along the Bouzidi guidelines, we write down the boundary magnetic vector-valued distributions $\boldsymbol{g}_i$ as
\begin{eqnarray}
\boldsymbol{g}_{\overline{i},\alpha}(\boldsymbol{x}_f, t + 1) &=&  2(\Delta -  1)\tilde{\boldsymbol{g}}_{i,\alpha}(\boldsymbol{x}_f,t) - \bigg ( \frac{(2 \Delta - 1)^2}{2 \Delta + 1} \bigg )\tilde{\boldsymbol{g}}_{i,\alpha}(\boldsymbol{x}_{ff},t) +  
\nonumber \\
&+& 2 \bigg ( \frac{2 \Delta -1}{2 \Delta + 1} \bigg ) \tilde{\boldsymbol{g}}_{\overline{i},\alpha}(\boldsymbol{x}_f,t) + \frac{1}{3} \bigg ( \frac{3 - 2 \Delta}{2 \Delta + 1} \bigg ) b_\alpha |_{wall} \ . \
    \label{boundarycondition_gmag}
\end{eqnarray}
A detailed formal analysis of the above BC prescription is straightforward and lengthy, so we skip it, for the purpose of a more objective exposition.

We emphasize that our considerations are related to the case of MHD flows bounded by insulating walls, 
where, due to the complete absence of induced currents, magnetic perturbations vanish. We have, thus, $\mathbf{b} = 0$ at the insulating surfaces, which is the magnetic analog of the no-slip boundary condition for the velocity field -- there is no contribution associated to the last term on the RHS of Eq. (\ref{boundarycondition_gmag}).


\section{\label{sim_res_sec} SIMULATION RESULTS}

To benchmark the CM-MRT framework for the lattice-Boltzmann simulations of MHD, we compare its numerical performances with some known exact solution of the MHD equations. We pick up, for this validation task, the Gold problem \cite{gold_1962}, which discusses a laminar pipe flow subject to a uniform transverse magnetic field.
\vspace{0.3cm}

\leftline{\it{Pipe Flow in the Presence of a Transverse Uniform Magnetic Field}}
\vspace{0.3cm}

\begin{figure}[b]
    \centering
    \includegraphics[width = 10cm]{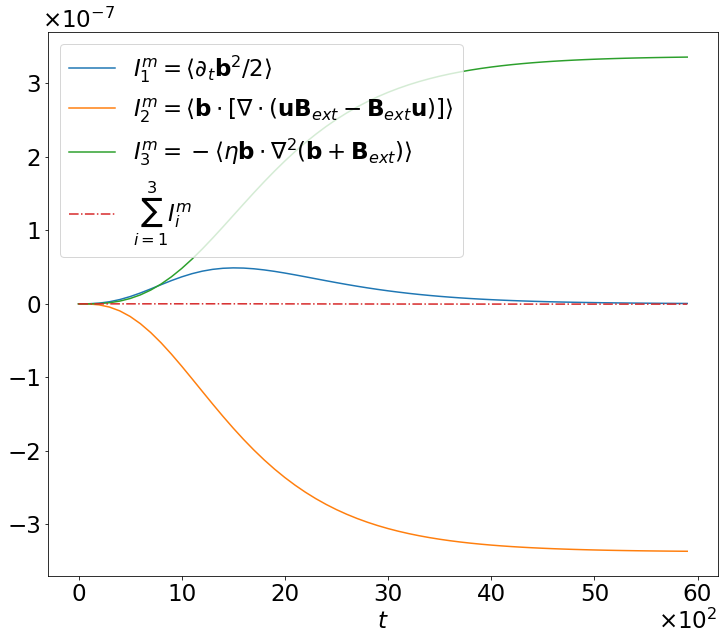}
    \caption{Magnetic energy balance for Ha = 5 and $R_m = 39$.}
    \label{conservationofmagneticenergy}
\end{figure}

Let the $z$ direction be identified with the pipe symmetry axis. The pipe's radius is $R$, its walls are insulators, and the external applied magnetic field is $\mathbf{B} = B_0\hat{y}$. The flow is driven by a constant pressure gradient $h = \partial p / \partial z$. Gold's exact laminar solution of Eqs. (\ref{NSMagForce}- \ref{completeIE}) gives $u_x=u_y=0$, $b_x=b_y=0$, and, in cylindrical coordinates,
\begin{eqnarray}
&&u_z(r,\theta) = \frac{R^2 h}{2 {\hbox{Ha}} \nu} \left [ \sum_{n = -\infty}^{\infty} \{e^{-\alpha r cos(\theta)} + (-1)^n e^{\alpha r cos (\theta)} \}\frac{I'_{n}(\alpha)}{I_n(\alpha)}I_n(\alpha r) e^{in\theta} \right ] \ , \  \\
&&b_z(r,\theta) = B_z(r,\theta) = \label{goldu} \\
&&= \frac{R^2 h}{2 {\hbox{Ha}} (\eta \nu)^{1/2}} \left [ \sum_{n = -\infty}^{\infty} \{ e^{-\alpha r cos (\theta)} - (-1)^n e^{\alpha r cos (\theta)} \} \frac{I'_{n}(\alpha)}{I_n(\alpha)}I_n(\alpha r) e^{in\theta} - 2r cos (\theta) \right ]
\label{goldb}
\ , \
\end{eqnarray}
where ${\hbox{Ha}} \equiv B_0L/\sqrt{\rho_0 \eta \nu}$ is the Hartmann number, a dimensionless parameter which estimates the ratio between magnetic and viscous forces \cite{shercliffbook}, and $I_n$ is the $n$-th order modified Bessel function.

The lattice size of our simulations is $n_x \times n_y \times n_z = 40 \times 40 \times 5$. Periodic boundary conditions are imposed along the $z$ direction (inlet and outlet velocities are the same). In lattice-Boltzmann units, the viscosity and the initial uniform velocity and magnetic fields are defined, respectively, as $\nu = 0.04$, $\mathbf{u}_0 = 0.08 \hat z$, and $\mathbf{b}=0$ (the Reynolds number is {\hbox{Re $\approx 40$}}). The explored Hartmann numbers, the pipe radius and the pressure gradient are, respectively, $0 \leq {\hbox{Ha}} \leq 15$, $R = (ny-1)/2 = 19.5$, and $h = u_0 \nu {\hbox{Ha}}/R^2$. Simulations were performed for different magnetic Reynolds numbers in the range $0.2 \leq R_m \leq 39$.

\begin{figure}[t]
    \begin{center}
    \includegraphics[width = 10cm]{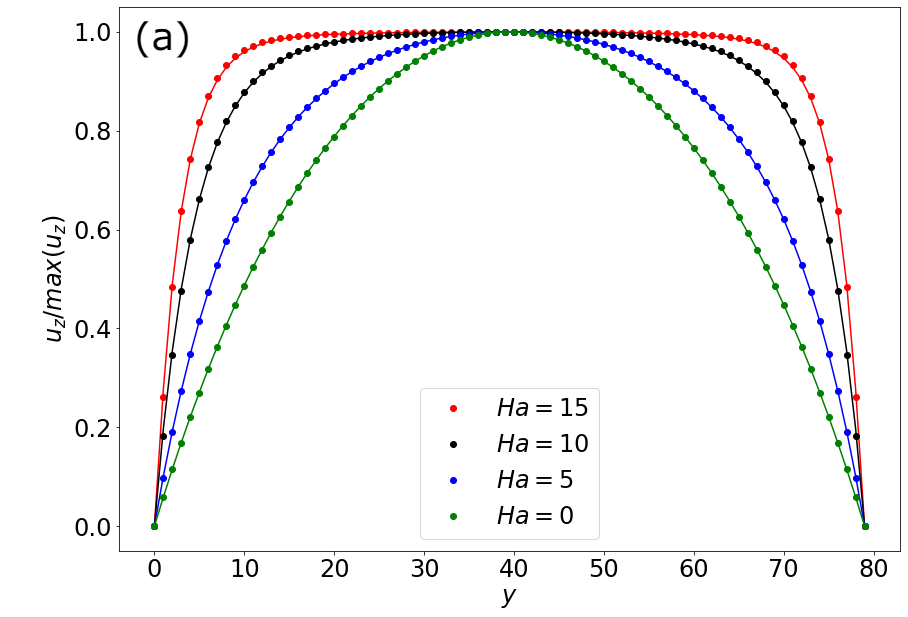}
    \includegraphics[width = 10cm]{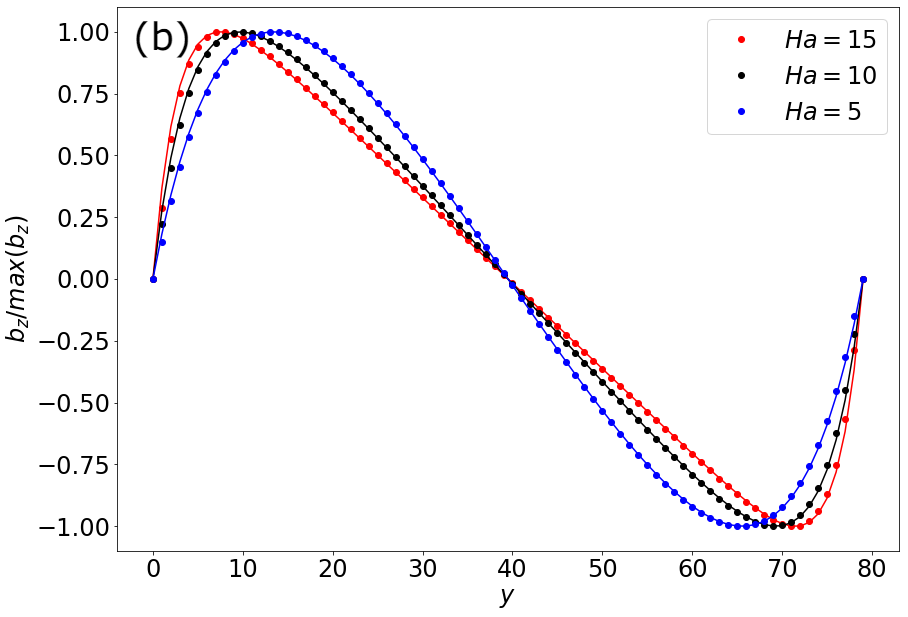}
    \caption{Results for simulations at $R_m = 15$. (a) Streamwise velocity field $u_z$, normalized by its centerline value. (b) The induced magnetic field $b_z$. Solid lines represent the Gold solution (\ref{goldu}) and (\ref{goldb}).}
    \label{magnetic_field_gold}
    \end{center}
\end{figure}

The magnetic energy balance can be readily derived from the scalar product of $\mathbf{b}$ with Eq. (\ref{IELBM}). The time dependence of each of the several energy production terms is shown in Fig. 1 for a set of magnetic Reynolds numbers at Ha = 5. It is clear from the data that the smaller is the magnetic Reynolds number, the stronger is the damping of $\partial_t \mathbf{b}^2$. We have been able to perform simulations down to $R_m = 0.2$, when instabilities start to occur because of the high value of the magnetic relaxation time $\tau_m \approx 40$. At the same grid resolution, in comparison, the usual magnetic BGK collision model loses its numerical stability at $\tau_m \geq 1$, which is related, in this scenario, to the minimum accountable magnetic Reynolds number of $R_m \approx 10$. Also, Fig. 2 clearly indicates that $\partial_t \mathbf{b}^2/2$ is accurately given by the summation of the several magnetic energy production terms (for illustration purposes, we show only the case of the highest magnetic Reynolds number ($R_m = 39$), where the energy production terms exhibit larger time dependent variations).

\begin{figure}[b]
    \centering
    \includegraphics[width = 16cm]{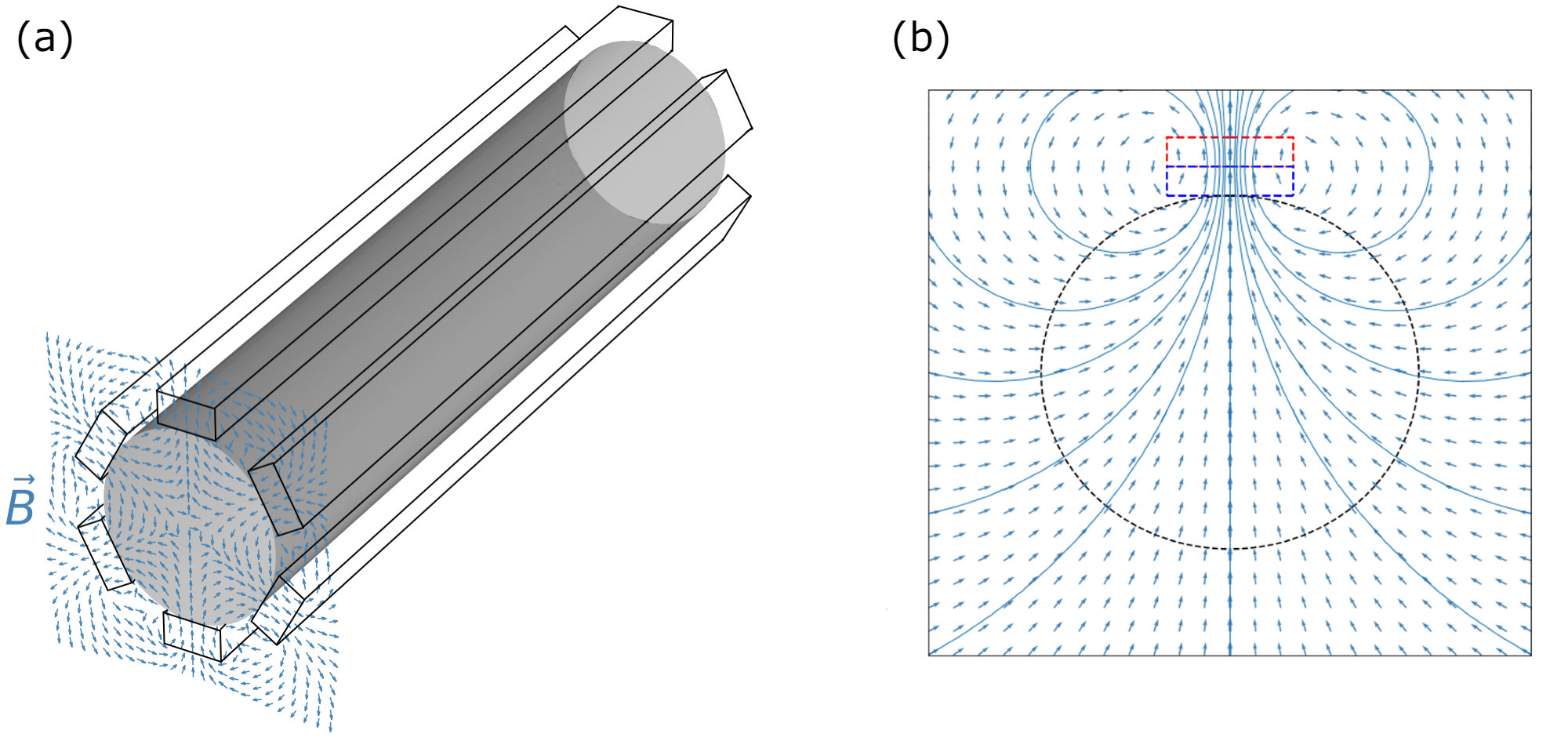}
    \caption{(a) Schematics of the pipe flow setup, with a representation of the positions of the six covering magnetic slabs. (b) The non-uniform magnetic field produced by just one of the magnetic slabs is represented in a cross-sectional plane, as modeled by Eqs. (\ref{Bslab1}) and (\ref{Bslab2}).}
    \label{non-uniform-magnetic-field}
\end{figure}

We have performed validation simulations for larger Hartmann numbers as well. Excellent convergence is attained for a better resolved lattice (which is necessary due to the existence of higher velocity gradients at larger Ha), with dimensions $n_x \times n_y \times n_z =  80 \times 80 \times 5$ and magnetic Reynolds number as low as $R_m = 15$ ($\tau_m = 4.5$). The results for the velocity and magnetic induced fields are shown in Fig.~3, in fine agreement with the exact solutions (\ref{goldu}) and (\ref{goldb}).
\vspace{0.3cm}


\leftline{\it{Pipe Flow in the Presence of a non-Uniform Magnetic Field}}
\vspace{0.3cm}

In order to discuss the feasibility of the MHD-LBM in a more complex geometric setting, we now carry out simulations for the velocity and magnetic induction fields in a laminar pipe flow surrounded by a hexagonal regular distribution of (ideally infinite) magnetic slabs, as depicted in Fig. 4a. The magnets have their north and south poles faces displaced in an alternate manner around the pipe. 

\begin{figure}[b]
    \centering
    \includegraphics[width = 16cm]{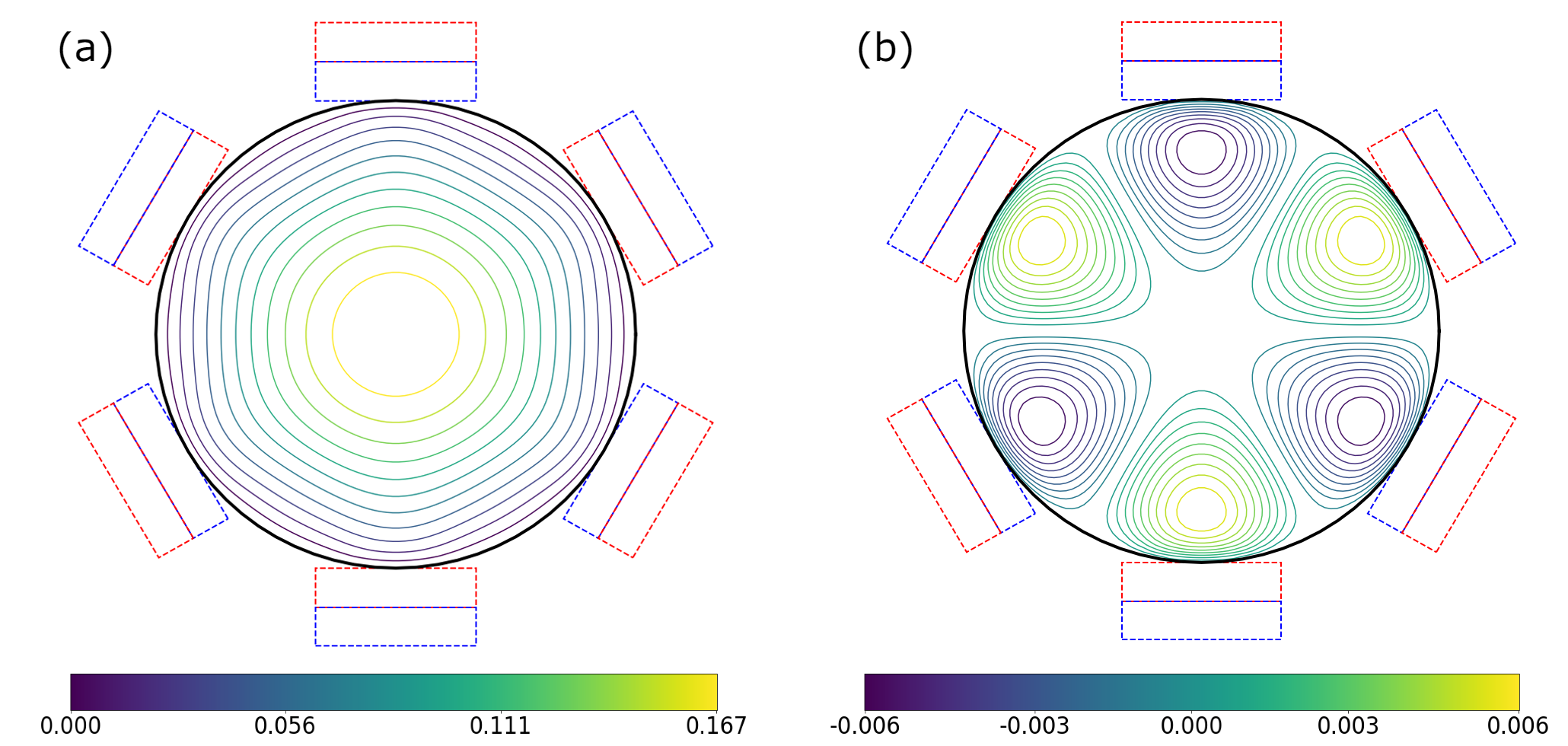}
    \caption{Level curves of (a) the velocity field and (b) the induced magnetic field. Both of them are parallel to the pipe's symmetry axis. The color bars indicate the values of the velocity and magnetic fields.}
    \label{contourmagnetic}
\end{figure}

\begin{figure}[t]
    \centering
    \includegraphics[width = 10cm]{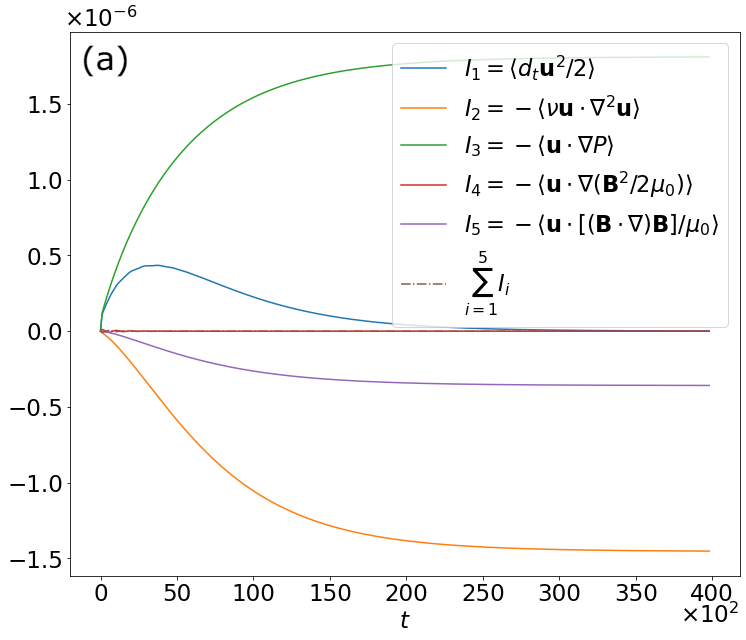}
    \centering
    \includegraphics[width = 10cm]{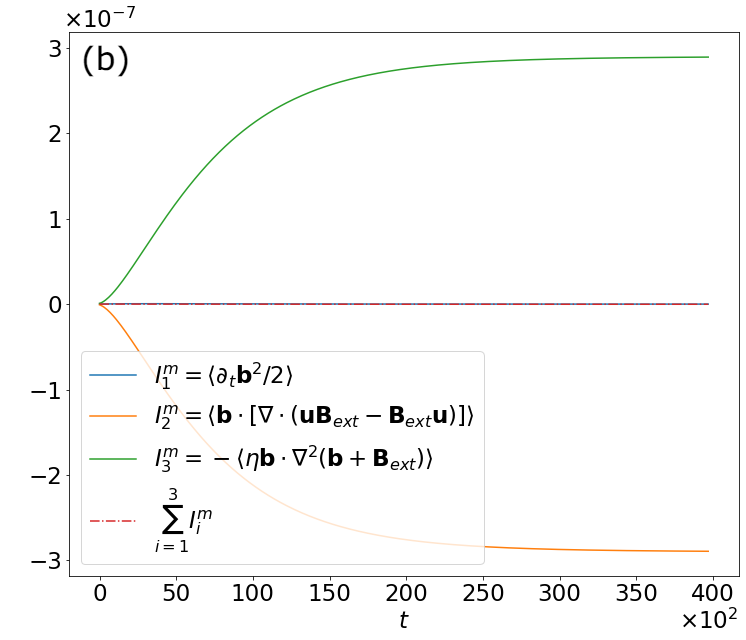}
    \caption{Energy balance analysis for (a) the Navier-Stokes
   Eq.  (\ref{NSMagForce}) and (b) the magnetic induction Eq. (\ref{IELBM}).}
    \label{magneticEnergy}
\end{figure}

Figure 4b illustrates the magnetic field lines produced by a magnetic slab which have its south pole oriented towards the pipe's interior. The expression for the magnetic field produced by this single slab is a straightforward exercise in magnetostatics \cite{Griffithsbook}. We have $B_z = 0$ due to symmetry and, in convenient units,
\begin{eqnarray}
B_x(x,y) &=& \ln \left [ \frac{(x+L)^2 + y^2}{(x-L)^2 + y^2} \right ] \ , \ \label{Bslab1} \\
B_y(x,y) &=& 2\left ( \arctan{\frac{x-L}{y}} - \arctan{ \frac{x+L}{y}  } \right ) \ , \ \label{Bslab2}
\end{eqnarray}
where $L$ is the width of the slab's rectangular cross section, which we postulate to have aspect ratio 2. We take, more specifically, $L=R/6$. The total external magnetic field applied to the pipe is the superposition of fields given by rotations of (\ref{Bslab1}) and (\ref{Bslab2}) in the $xy$ plane.

\begin{figure}[h]
    \centering
    \includegraphics[width = 10cm]{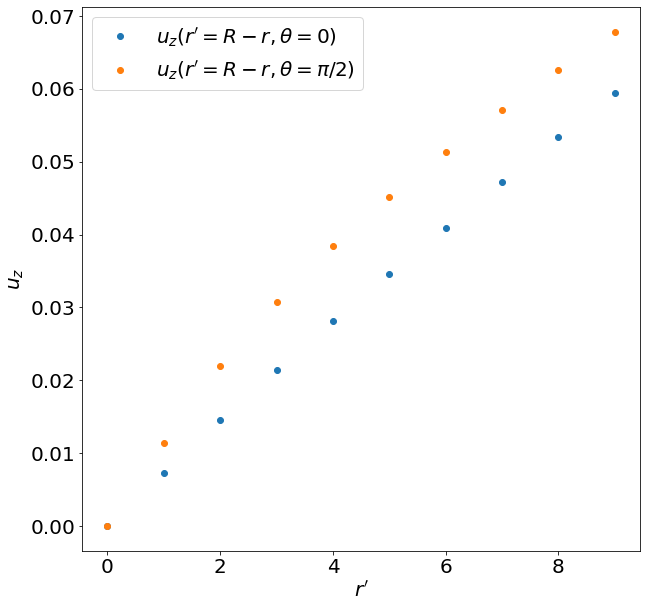}
    \caption{Near-wall velocity profiles along the $\theta = 0$ and $\theta = \pi/2$ directions.}
    \label{velocities_nonuniform}
\end{figure}

Our LBM simulations were performed on a lattice with dimensions  $n_x \times n_y \times n_z =  80 \times 80 \times 5$ 
at Re = 80 and $R_m = 3.16$. Using the maximum value of the external magnetic field on the pipe's surface to define a {\it{proxy}} Hartmann number, we obtain Ha $\approx 10$. Since there is no available exact solution for this particular pipe flow case, we focus on the symmetry properties of the numerical solutions and the accuracy of the evaluated kinetic and magnetic energy balances.

    


As the Lorentz force (\ref{lforce}) is symmetric under the mapping $\mathbf{B} \to -\mathbf{B}$, the solution for the velocity field $\mathbf{u} = u(x,y) \hat z$ must be invariant under rotations of $\pi/3$ around the $z$ axis, which is clearly indicated from the velocity level curves of $u(x,y)$ provided in Fig. 5a. The magnetic induction equation (\ref{IELBM}), on its turn, implies that $\mathbf{b}$ is anti-symmetric upon the substitution of $\mathbf{B}_{ext}$ by $-\mathbf{B}_{ext}$. This is actually verified from the magnetic level curves given in Fig. 5b, in combination with discrete rotations by $\pi/3$.

The kinetic and magnetic energy balances are reported in Fig. 6. As we see, the herein addressed CM-MRT lattice-Boltzmann simulations respect energy conservation all the way along the flow dynamic evolution, up to the asymptotic stationary regime.

We scrutinize, furthermore, the near-wall velocity profiles for $\theta = \pi/2$ (maximum intensity for the normal magnetic field) and $\theta = 0$ (minimum intensity for the normal magnetic field). It turns out, as it can be inferred from Fig. 7, that the velocity varies by a non-negligible amount up to $r'/R \approx 0.2$, when these two angular directions are compared, a result that is somewhat surprising, given that the magnetic field is noticed to decay in a fast way inside the pipe and the proxy Hartmann number is not very high. 

Induced local effects are thus relevant and sensitive to the intensity and orientation of the magnetic field in the near-wall region. We point out that the reorganization of the shearing distribution around the pipe can have interesting consequences in connection with the phenomenon of drag reduction, usually investigated for flows subject to uniform magnetic fields at much higher Hartmann's numbers \cite{Tsinober,moriconi2020,chaudhary_vanka_thomas2010,zikanov_krasnov_boeck_thess_rossi,krasnov_zikanov_boeck_2012}. It is important to note, however, that once the magnetic field considerably drops in the bulk of the flow in our specific setting, the velocity profile near its centerline is essentially axisymmetric and locally parabolic.

\section{\label{conc_sec} CONCLUSIONS}

We have introduced an improved realization of the LBM for the simulation of MHD flows, which overcomes the severe instability problems associated with the usual simpler treatment of magnetic relaxation \cite{DELLAR2002}. The central issue we have focused on is that the straightforward BGK modeling of magnetic diffusion, based on a single relaxation time scale, although not conceptually or technically mistaken, turns out to be computationally demanding and, in practice, of little usefulness in the limit of very low magnetic Reynolds numbers, where the need of a much finer lattice resolution becomes inexorable, even for laminar flows.

Our alternative lattice-Boltzmann approach to MHD consists in postulating a set of magnetic relaxation time scales, in close analogy with what has been already carried in the pure hydrodynamic \cite{derosis_huang_coreixas2019} and thermal transport \cite{YOSHIDA2010} contexts. We also have, similarly, worked out a general procedure for the imposition of the Dirichlet magnetic boundary conditions in the case of insulating walls, motivated by the well-established Bouzidi interpolation schemes \cite{bouzidi2001,li_mei_klausner2013}. The central moment - MRT LBM so devised for MHD leads to accurate comparisons with the analytic solution of the Gold's problem for a laminar pipe flow subject to a transverse uniform magnetic field. 

Therefore, confident on the method's validity, we have explored as a case study the more complex situation of a laminar pipe flow subject to the presence of external non-homogeneous magnetic fields with a sixfold symmetry. We then find that the balance equations for the production of kinetic and magnetic energy are finely satisfied. Furthermore, we note, as a relevant remark, that lattice effects associated to the underlying cubic lattice do not spoil the expected symmetry properties of the flow.

The present study opens the way for the performance of computationally efficient simulations of MHD in turbulent regimes, still in the challenging domain of lower magnetic Reynolds numbers. Further improvements and generalizations are in order, as the definition of alternative wall boundary conditions for the induced magnetic field.
\vspace{0.5cm}

\leftline{\it{Acknowledgments}}
\vspace{0.3cm}

This work was partially supported by the Conselho Nacional de Desenvolvimento Científico
e Tecnológico (CNPq), Brazil, and Petrobras (COPPETEC project number 21389).

\nocite{*}

\end{document}